\documentclass[12pt,preprint]{aastex}

\begin{document}
\title{Properties of Voids in the 2dFGRS Galaxy Survey}
\author{Anton V.\ Tikhonov}
\affil{ St. Petersburg State University,St.\ Petersburg, Russia}
\email{avt@gtn.ru, ti@hotbox.ru}
\begin{abstract}

A method for detecting voids in the galaxy distribution is
presented. Using this method, we have identified 732 voids with a
radius of the seed sphere $R_{seed}> 4.0h^{-1}$~Mpc in a
volume-limited sample of galaxies from the southern part of the
2dFGRS survey. 110 voids with $R_{seed}> 9.0h^{-1}$~Mpc have a
positive significance. The mean volume of such voids is $19 \cdot
10^3 h^{-3}$~Mpc$^3$. Voids with $R_{seed}> 9.0h^{-1}$~Mpc occupy
55\% of the sample volume. We construct a dependence of the
volumes of all the identified voids on their ranks and determine
parameters of the galaxy distribution. The dependence of the
volume of voids on their rank is consistent with a fractal model
(Zipf's power law) of the galaxy distribution with a fractal
dimension $D \approx 2.1$ (given the uncertainty in determining
the dimension using our method and the results of a correlation
analysis) up to scales of $25h^{-1}$~Mpc with the subsequent
transition to homogeneity. The directions of the greatest
elongations of voids and their ellipticities (oblateness) are
determined from the parameters of equivalent ellipsoids. The
directions of the greatest void elongations have an enhanced
concentration to the directions perpendicular to the line of
sight.

\end{abstract}


DOI: 10.1134/S1063773706110028


\section{Introduction}
Information about the structuring of the spatial distribution of
luminous matter (the form and degree of clustering of galaxies and
clusters of galaxies, the characteristic scales of clustering,
etc.) characterizes the physical conditions at the formation epoch
of the observed inhomogeneities.

The first galaxy redshift surveys revealed that the spatial
distribution of galaxies is essentially inhomogeneous: they are
gathered in groups and clusters, which, in turn, form a complex
network of filaments and walls that are threaded by tunnels and
that bound voids (a "foamy" or "cellular" structure).

In the early 1980s, it became clear that giant voids in the
distribution of galaxies are common in the observable Universe
(Einasto et al. 1980; Kirshner et al. 1981), but their
significance was understood later. The history of void detection
and description was presented by Rood (1988).

Describing the distribution and parameters of voids in the
distribution of galaxies is important for understanding the
formation conditions of the large-scale structure of the
observable Universe and imposes constraints on cosmological model
parameters (Peebles 2001; Berlind and Weinberg 2002). CDM models
predict a certain amount of matter and, hence, galaxies in voids.
However, studies have shown that dwarf galaxies are located in the
same structures as normal galaxies (Gottlober et al. 2003; Hoyle
and Vogeley (2002) and references therein).

The model of cellular structure allows a void to be defined as a
connected region of space with a reduced density of galaxies
compared to the regions containing the formations that constitute
the large-scale structure (walls, filaments, and clusters) or
completely free from galaxies. The methods for analyzing the
distributions of void parameters complement the methods for
studying the distribution of galaxies; in particular, describing
the spectrum of void volumes allows the morphology of the
distribution under study to be determined. VPF (Void Probability
Function) and UPF (Underdensity Probability Function) are the
standard statistical methods that determine high order correlation
functions. VPF determines the probability that a sphere of radius
$r$ placed in a given distribution will be empty; UPF measures the
frequency of spheres with a density contrast $\delta\rho/\rho$
below a certain threshold.

There are several algorithms for identifying voids in a point
distribution that do not determine the void shapes a priori. The
most developed algorithms are given in the papers by El-Ad and
Piran (1997) -- the method of nonoverlapping spheres, Aikio and
Mahonen (1998) -- the method of maxima of the field of distances,
Gaite (2005) -- the method of Delaunay and Voronoi tessellations,
and Shandarin et al. (2006) -- using a smoothed density field.
Different algorithms provide different lists of voids.

The shapes of voids along with the spectrum of their sizes are of
considerable interest. For example, Plionis and Basilakos (2002)
analyzed the distributions of void sizes and shapes in the PSCz
survey and compared them with artificial distributions obtained
using various CDM models.

Hoyle and Vogeley (2004) have already searched the 2dFGRS survey
for voids using an algorithm similar to that presented by El-Ad
and Piran (1997) and Croton et al. (2004) based on VPF. Using the
method of searching for maximum spheres free from galaxies with
masses (luminosities) below a certain value, Patiri et al. (2005)
determined the statistical parameters of such voids in the
distribution of 2dFGRS galaxies.

\section{THE METHOD OF SEARCHING FOR VOIDS}
The void finder algorithm presented here is similar to the
procedure described by El-Ad and Piran (1997) and differs from the
approach used by Hoyle and Vogeley (2004.

Here, we searched for voids that were completely free from
galaxies (without separating the galaxies into wall and void
galaxies), because the procedure for constructing a volume-limited
sample leaves only fairly bright galaxies that "outline" the
structure (formed inside massive halos) in the sample.

Our method uses a standard (for a number of related methods) ad
hoc parameter (in our case, the parameter $k$) that defines the
size of the tunnel (hole) in the distribution of galaxies into
which a given void penetrates during the construction. The method
used by Shandarin et al. (2006) is free from such parameters.
However, when a smoothed density field is constructed, the void
parameters depend on the smoothing length (e.g., in the case of
smoothing by a Gaussian filter) and on the threshold density that
separates the regions with enhanced and reduced densities.
Besides, this approach yields very irregular voids.

The method presented here is fairly simple, flexible, and suitable
for searching for voids defined as regions completely free from
galaxies.

An orthogonal three-dimensional grid is constructed in the volume
of the sample under study to attribute the grid points to a
particular void. The voids are identified from large to small
ones. First, the sphere with the largest possible radius that fits
into the empty (free from galaxies) regions of the galaxy
distribution and the geometrical boundaries of the sample is
identified. Since the voids are assumed to lie strictly within the
geometrical boundaries of the sample, the radius for a given grid
point is defined as the minimum of the smallest distance from the
given grid point to galaxies and the minimum distance from the
point to the boundary.

Inside the sample volume, galaxy-free seed spheres are
sequentially searched for (first, the largest sphere is searched
for) followed by their expansion through the addition of the
spheres whose centers are located within the already fixed part of
the void and whose radii $R_{sph}$ are no less than the radius of
the seed sphere multiplied by the coefficient $k = 0.9$ ($R_{sph}
> 0.9 \cdot R_{seed}$, where $R_{seed}$ is the radius of the seed
sphere). Thus, the spheres added to the void overlap with the
region already attributed to the void by more than 30\% of their
volume. Subsequently, the empty sphere with the largest possible
radius is identified again by taking into account the void
identified at the previous step (all the grid points attributed to
this void are excluded from the analysis), this sphere is then
expanded, and so on. The identification of voids continues as long
as $R_{seed}$ is larger than a certain value (in our case, $4.0
h^{-1}$~Mpc). The radius of the sphere with the volume per galaxy
in the volume-limited under study (see below) is $R_{gal} \approx
5.0 h^{-1}$~Mpc.

At $k = 0.9$, the voids "outline" fairly well the regions between
galaxies and, at the same time, retain a regular shape, which is
convenient when voids are approximated by ellipsoids. At $k = 1$,
the voids are strictly spherical. If we decrease $k$, then the
voids begin to penetrate into increasingly small holes and the
void shapes become increasingly irregular. At small $k$, one (the
first) void fills most of the sample volume. The coefficient $k =
0.9$ used here is a compromise chosen after the void construction
in point distributions with different properties. The voids
constructed in this way (at $k = 0.9$) are identified in
arbitrarily shaped volumes. On the other hand, the voids are found
to be separated from one another and to be fairly "thick"
throughout the volume, which allows them to be approximated by
triaxial ellipsoids.

We divided the identified voids into the voids that completely lie
within the geometrical boundaries of the sample and the voids that
touch the sample boundaries; hence, their volumes are
underestimated and their shapes are distorted by the boundary
effect.

\section{THE DATA}
The second release of the 2dFGRS galaxy and quasar survey (Colless
et al. 2001, 2003) includes $\sim250 000$ galaxies and up to 25
000 quasars. A multiobject spectrograph with a $2^\circ$ field of
view capable of taking 400 spectra in a single exposure was used
to determine the redshifts. The APM ($b_j < 20.5$) photometric
catalog formed the basis for the survey. All of the galaxies with
Galactic extinction-corrected apparent magnitudes $b_j$ in the
range $14.5 < b_j < 19.5$ were selected for the 2dFGRS survey. The
main region in the sky covered by the 2dFGRS survey consists of
two strips: $80^\circ$ in right ascension and $15^\circ$ in
declination near the South Galactic Pole and $75^\circ$ in right
ascension and $10^\circ$ in declination in the Northern Galactic
Hemisphere. Additionally, galaxies in 99 randomly located fields
in the high-latitude part of the South Galactic Hemisphere were
imaged for the survey. The survey covers $\sim2000$ square degrees
and has a median depth of $z=0.11$ for galaxies with $b_j <
19.45$. In this paper, we use a sample from the southern part of
2dFGRS and consider galaxies with a redshift quality parameter
($z$) $quality \ge 3$. The boundaries of the sample are: $21.84^h
< \alpha < 23.44^h$, $-35^\circ < \delta < -24^\circ$, and $0.02 <
z < 0.121$. At $z_{max}=0.121$, the volume-limited sample contains
the largest number (7219) of galaxies for these angle boundaries.
In such a sample, all galaxies with absolute magnitudes $M_{abs} <
-19.92$ are observable from any point of the sample with apparent
magnitudes $b_j < 19.2$. To estimate the absolute magnitudes of
galaxies, we calculated the combined evolutionary correction
$e(z)$ and the correction for the shift of the range of the
galaxy's radiation to the longer wavelengths (the $K(z)$
correction) used to calculate the luminosity function for 2dFGRS
galaxies (Norberg et al. 2002): $K(z)+e(z) = (z+6 \cdot z^2)/(1+20
\cdot z^3)$. To determine the metric distances from the redshift
using a formula given, for example, in Hogg et al. (1999), we used
the Hubble parameter $H_0 = 65$ km s$^{-1}$Mpc$^{-1}$ ($h=H/H_0$,
where $H$ is the true value of the Hubble constant) and the
density parameters $\Omega_{\Lambda} = 0.7$, $\Omega_0 = 0.3$. In
such a cosmology, $z_{max}=0.121$ corresponds to a distance of
$542.54h^{-1}$~Mpc.

\section{IDENTIFIED VOIDS}
We identified a total of 732 voids with $R_{seed}> 4h^{-1}$~Mpc in
the work sample using the algorithm described above.

The configuration of voids and their volumes to some extent depend
on the chosen step of the three dimensional grid. In the presented
implementation of the algorithm, we chose the grid step to be
$Step \approx 1.0$~Mpc. The void volumes were estimated as $Step^3
\times N$, where $N$ is the number of grid points inside a given
void. The volume per galaxy in our sample is $513h^{-3}$ Mpc$^3$.
The largest found void occupies a volume of $85595h^{-3}$ Mpc$^3$.
The radius of the first (largest) seed (initial) sphere is
$R_{seed}^1=21.3h^{-1}$~Mpc. In Fig.1, the void volume is plotted
against the radii $R_{seed}$ of the seed spheres. As expected, the
scatter of volumes increases with $R_{seed}$.

We determined the significance of voids by comparing the
identified voids with the list of voids obtained using the same
algorithm in a random sample with the same mean density. The
significance is $P(r) = 1 - N_{random}(r)/N_{survey}(r)$, where
$N_{survey}(r)$ is the number of seed spheres of voids with radii
$R_{seed}>r$ identified in the 2dFGRS sample and $N_{random}(r)$
is the number of seed spheres identified in a homogeneous
distribution within the same boundaries and with the same mean
density as that for the galaxy sample (Fig. 2). 110 voids of the
2dFGRS sample with radii $R_{seed}> 9h^{-1}$~Mpc have a positive
significance; 23 of them do not touch the boundaries. All the
significant voids touch at least one galaxy in the process of
construction. Figure 3 shows the projection of the distribution of
galaxies with $-32^\circ < \delta < -27^\circ$ and the centers of
110 voids with $R_{seed}> 9h^{-1}$~Mpc.

The mean volume of the voids with significance $R_{seed}>
9h^{-1}$~Mpc is $18666 h^{-3}$~Mpc$^3$. The ratio of the total
volume of the seed spheres to the total volume of the voids
($R_{seed}> 9h^{-1}$~Mpc) is 0.31. On average, the seed sphere
occupies 0.32 of the volume of the void "grown" from it for
significant voids. Figure 4 shows the distribution of void
volumes. A nearly power-law decrease in the number of voids with
increasing volume is observed.

The voids identified using the algorithm presented here with
$R_{seed}> 9h^{-1}$~Mpc occupy 55\% of the sample volume.

\section{VOLUME STATISTICS. PARAMETERS OF THE GALAXY DISTRIBUTION}
The distribution of galaxies (the behavior of the density with
distance) is known to obey a power law on small scales. This is
occasionally interpreted as fractality, self-similarity in a
certain interval of scales (up to $20-30h^{-1}$~Mpc). Fractal
methods yield a fractal dimension of $\sim 2$ (Coleman and
Pietronero 1992; Baryshev and Teerikopi 2005; Tikhonov 2005).
Gaite and Manrubia (2002) showed that the fractal dimension could
be derived from the distribution of void volumes in a fractal
structure: Zipf's power-law (Zipfs 1949) dependence of the volume
on rank (the largest void has rank 1, the next void has rank 2,
etc.) is valid for the volumes of the voids identified in a
three-dimensional fractal distribution: $V(Rank) \propto
Rank^{-z_f}$, $z_f = 3/D$, where $D$ is the fractal dimension of
the distribution.

Gaite and Manrubia (2002) found no power-law dependence using some
of the void catalogs and suggested that Zipf's law was not
observed in the cases considered because of the algorithms used
for void identification. Artificial fractal distributions,
including the distribution obtained using the "Cantor dust"
algorithm also known as a $\beta$-cascade (Paladin and Vulpiani
1987), revealed that the algorithm presented in this paper
overestimates the fractal dimension $D$ by $0.1-0.2$ at $D \ge 2$
and poorly estimates it at $D < 2$ ($D > 2$ is a fundamental
limitation for estimating the dimension using Zipf's law (Gaite
2006)); a well-defined power-law dependence is observed in all
cases. In Fig.5, the volumes of all 732 voids identified in the
2dFGRS sample is plotted against their ranks. The bend of the
dependence at the highest ranks (smallest volumes) is due to the
limitation on the radius of the seed sphere from below. The
dependence in Fig.5 can be interpreted in terms of Zipf's law from
small volumes to $V_{break} \sim 10^4 h^{-3}$~Mpc$^3$. After the
smooth break, the slope of the dependence toward the larger
volumes (lower ranks) becomes less than unity, which rules out the
fractal interpretation (Gaite and Manrubia 2002). Gaite (2005)
interpreted this behavior of the dependence as a manifestation of
the transition to homogeneity in the given distribution. The
effective diameter (the diameter of a sphere with the same volume)
for $V_{break}$ is $\sim26h^{-1}$~Mpc. This diameter virtually
coincides with the scale on which the correlation gamma-function
deviates significantly from a power law (Tikhonov et al. 2000;
Tikhonov 2005). The slope before the break (at high ranks) is $z_f
\approx 1.4$, which corresponds to a fractal dimension of $D=3/z_f
\approx 2.14$ and also agrees with the results of the correlation
analysis (e.g., of the 2dFGRS survey (Tikhonov 2005)).

\section{VOID SHAPES}
Once we compiled the list of voids (attributed three-dimensional
grid points to a particular void), we determined the void centers
and calculated the moments of inertia of the bodies formed by the
voids. We analyzed the void shapes using the parameters of their
equivalent ellipsoids. We constructed the $3\times3$ matrix of
moments of inertia $I_{ij}$ and found its eigenvalues $\lambda_i$,
which are equal to the principal moments of inertia, from the
condition $det(I_{ij}-\lambda \cdot E)=0$ (where $E$ is a
$3\times3$ unit matrix); these principal moments of inertia were
used to determine the semiaxes of the equivalent triaxial
ellipsoid. The eigenvectors of the matrix $I_{ij}$ give the
directions of the semiaxes. The direction of the greatest void
elongation coincides with the direction of the major axis of the
equivalent ellipsoid. The distribution of the directions of void
elongations is not quite homogeneous (Fig. 6): we can note a
crowding toward the directions perpendicular to the line of sight
(especially of the major axes of the voids that do not touch the
boundaries). No void elongation along the line of sight is
observed. The ellipticity (oblateness) of voids was defined as
$\epsilon = 1 - c/a$, where $a$ and $c$ are the semimajor and
semiminor axes, respectively. 37 and 8 voids with $R_{seed} >
9.0h^{-1}$~ Mpc have $\epsilon > 0.4$ and $\epsilon < 0.15$,
respectively; the largest ellipticity is  $\epsilon = 0.61$. The
voids are distributed in ellipticity rather homogeneously (Fig.
7).

\section{CONCLUSIONS}
In this paper, we presented an algorithm for identifying voids
without prior determination of their shapes. We identified 110
signi?cant voids with radii of the seed spheres $R_{seed} > 9.0
h^{-1}$~Mpc in the southern part of the volume-limited sample from
the 2dFGRS galaxy survey. These voids occupy 55\% of the sample
volume. The table gives parameters of the nine largest voids with
volumes larger than $40\cdot 10^3h^{-3}$~Mpc$^3$ (Fig. 8). The
mean effective radius of voids, $16.46h^{-1}$~Mpc, is considerably
smaller than that obtained by Hoyle and Vogeley (2004) — the
algorithm presented here identifies voids of smaller volumes,
because the galaxies are not divided into wall and void galaxies
(we used all of the galaxies included in the work sample). The
distribution of the directions of the major axes of voids is not
quite homogeneous. There is no tendency for them to be grouped
near the line-of-sight direction, but there is a crowding of the
directions of the major axes of voids near the direction
perpendicular to the line of sight; this can be observed in
redshift space if the voids contract homogeneously in the comoving
coordinates during the cosmological evolution (Ryden and Melott
1996). No systematic trends are observed in the distribution of
void ellipticities (oblateness). The dependence of the volumes of
voids on their ranks exhibits a power-law portion at high ranks
that can be interpreted in terms of Zipf's law with a fractal
dimension of $D \approx 2.1$ of the galaxy distribution. The
behavior of this dependence at lower ranks (at void volumes larger
than $10^4 h^{-3}$Mpc$^3$), which is indicative of the transition
to a homogeneous distribution from a scale of $\sim25h^{-1}$~Ìïê,
is consistent with the results of the correlation analysis of the
2dFGRS sample.

\section{ACKNOWLEDGMENTS}
I wish to thank D. Makarov and J. Gaite for helpful discussions
and the Administration of St.-Petersburg for support (grant
PD05-1.9-117).

\section{REFERENCES}
1. J. Aikio and P. Mahonen, Astrophys. J. 497, 534 (1998).

2. Y. Barishev and P. Teerikorpi, astro-ph/0505185 (2005).

3. A. Berlind and D. H. Weinberg, Astrophys. J. 575, 587 (2002).

4. P. H. Coleman and L. Pietronero, Phys. Rep. 213, 311 (1992).

5. M. Colles, G. Dalton, S.Maddox, et al. (The 2dFGRS Team),Mon.
Not. R. Astron. Soc. 328, 1039 (2001).

6. M. Colles, B. Peterson, C. Jackson, et al. (The 2dFGRS Team)
astro-ph/0306581 (2003).

7. D. J. Croton, M. Colles, E. Gaztanaga, et al., Mon. Not. R.
Astron. Soc. 352, 828 (2004); astro-ph/ 0401406 (2004).

8. J. Einasto, M. Joeveer, and E. Saar, Mon. Not. R. Astron. Soc.
193, 353 (1980).

9. H. El-Ad and T. Piran, Astrophys. J. 491, 421 (1997).

10. J. Gaite, astro-ph/0510328 (2005).

11. J. Gaite, private communication (2006).

12. J. Gaite and S. C. Manrubia, Mon. Not. R. Astron. Soc. 335,
977 (2002); astro-ph/0205188 (2002).

13. S. Gottlober, E. L. Locas, A. Klypin, and Y. Hoffman,
astro-ph/0305393 (2003).

14. D.W. Hogg, astro-ph/9905116 (1999).

15. F. Hoyle and M. S. Vogeley, Astrophys. J. 566, 641 (2002);
astro-ph/0109357.

16. F. Hoyle and M. S. Vogeley, Astrophys. J. 607, 751 (2004);
astro-ph/0312533.

17. R. P. Kirshner, A. Jr. Oemler, P. L. Schechter, and S. A.
Shectman, Astrophys. J. 248, L57 (1981).

18. P. Norberg, S. Cole, C. M. Baugh, et al., Mon. Not. R. Astron.
Soc. 336, 907 (2002); astro-ph/0111011 (2001).

19. G. Paladin and A. Vulpiani, Phys. Rep. 156, 147 (1987).

20. S. G. Patiri, J. Betancort-Rijo, F. Prada, et al., submitted
to Mon. Not. R. Astron. Soc.; astro-ph/ 0506668 (2005).

21. P. J. E. Peebles, Astrophys. J. 557, 495 (2001).

22. M. Plionis and S. Basilakos, Mon. Not. R. Astron. Soc. 330,
399 (2002).

23. H. J. Rood, Ann. Rev. Astron. Astrophys. 26, 245 (1988).

24. B. S. Ryden and A. L. Melott, Astrophys. J. 470, 160 (1996);
astro-ph/9510108.

25. A. V. Tikhonov, Pis'ma Astron. Zh. 31, 883 (2005) [Astron.
Lett. 31, 787 (2005)].

26. A. V. Tikhonov, D. I.Makarov, and A. I. Kopylov, Bull. Spec.
Aastrofiz.Obs., Russ. Akad. Sci. 50, 39 (2000); astro-ph/0106276
(2001).

27. S. Shandarin, H. A. Feldman, K. Heitmann, and S. Habib, Mon.
Not. R. Astron. Soc. 367, 1629 (2006).

28. G. K. Zipf, Human Behavior and the Principle of Least Effort
(Addison-Wesley, Massachusetts, 1949).

\newpage
\textwidth=16cm \oddsidemargin=-1.0cm

\begin{table}[!th]
\caption{Parameters and locations of the nine largest voids in the
2dFGRS survey \label{Tab1}}
\begin{tabular}{cccccccccccc}
\hline\hline $ N $& $R_{seed}$ & Volume &\multicolumn{3}{c}{Void
centers}&\multicolumn{3}{c}{Ellipsoid axes} &
$\epsilon$&\multicolumn{2}{c}{Direction a} \\
& & &r&$\alpha$&$\delta$&a&b&c& &$\alpha$&$\delta$\\
&(Mpc*) &(Mpc$^3$) & (Mpc)&(hours) &(deg.)&(Mpc)&(Mpc)&(Mpc)& &(hours)&(deg.) \\
\hline
        1**&21.33&68696     &494.325  & 22.684 &-27.966 &      27.87&   25.60&   23.08&    0.17 &  8.99 &   -82.6  \\
        2&20.19 &85595    &276.871 &23.032 &   -28.116 &      38.88&   24.43&   21.72&    0.44 &  9.21 &   -31.3   \\
        3**&19.99 &57727    &432.158& 22.622 & -30.581  &      27.77&   22.81&   21.85&    0.21 &  6.68 &   -76.2   \\
        4&19.25&66112     &463.606& 22.371  &  -27.317 &      34.49&   21.99&   20.95&    0.39 &  0.46 &    19.6  \\
        5&19.24 &59241    &510.441& 22.038 &  -26.597  &      27.26&   25.42&   20.57&    0.25 &  5.81 &   -50.5   \\
        6&19.05&44355     &499.752&22.147  &  -32.629  &      24.39&   21.30&   20.45&    0.16 & 11.74 &    16.5   \\
        7&18.96 &66518    &317.730& 22.465  & -28.069  &      31.83&   23.64&   21.45&    0.33 &  1.22 &   -55.0   \\
        8&18.45 &46601    &206.704& 22.351 &    -29.214 &     25.00&   22.96&   19.50&    0.22 &  7.91 &   -49.3   \\
        20&14.25 &41221   &448.997& 22.020 &   -26.783 &      27.99&   20.88&  17.41&    0.38&  11.52 &    45.0   \\
\hline
\end{tabular}
\vspace{1ex}

Note. N is the numbering that corresponds to the order of void
construction; the void centers are the equatorial coordinates
($\alpha_{2000}$, $\delta_{2000}$) of the void centers determined
as the centers of mass of the figures identified by the algorithm;
($\epsilon$) is the ellipticity; Direction $a$ is the direction of
the major axis a of the equivalent ellipsoid.

* --- In what follows, the units of $h^{-1}$ are used in the table.

** --- The voids do not touch the sample boundaries.
\end{table}

\newpage

\begin{figure}
\centerline{%\psfig{figure=f1.eps,height=10cm,angle=0}
\includegraphics[]{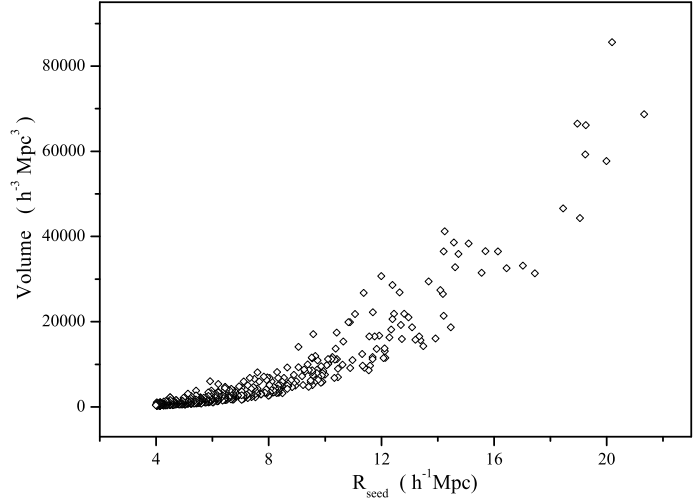}
} \figcaption{Distribution of void volumes vs. radii $R_{seed}$ of
the seed spheres of these voids.}
\end{figure}

\begin{figure}
\centerline{%\psfig{figure=f1.eps,height=10cm,angle=0}
\includegraphics[]{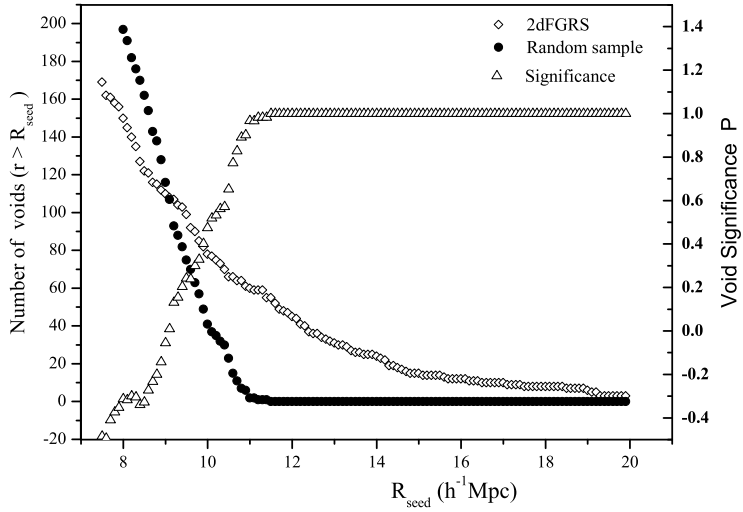}
} \figcaption{Number of voids with the radii of the seed spheres
larger than $R_{seed}$ in the 2dFGRS sample (diamonds) and the
random sample (circles). The triangles indicate the statistical
significance of voids $P$ as a function of $R_{seed}$.}
\end{figure}

\begin{figure}
\centerline{%\psfig{figure=f1.eps,height=10cm,angle=0}
\includegraphics[]{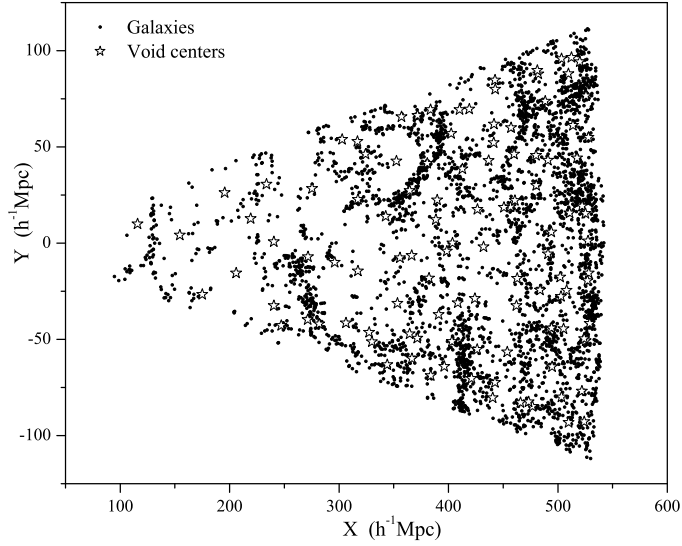}
} \figcaption{Projection of the distribution of the centers of 110
voids with $R_{seed}> 9h^{-1}$~Mpc (asterisks) and the
distribution of galaxies from the 2dFGRS work sample (dots) with
$-32^\circ < \delta < -27^\circ$. }
\end{figure}

\begin{figure}
\centerline{%\psfig{figure=f1.eps,height=10cm,angle=0}
\includegraphics[]{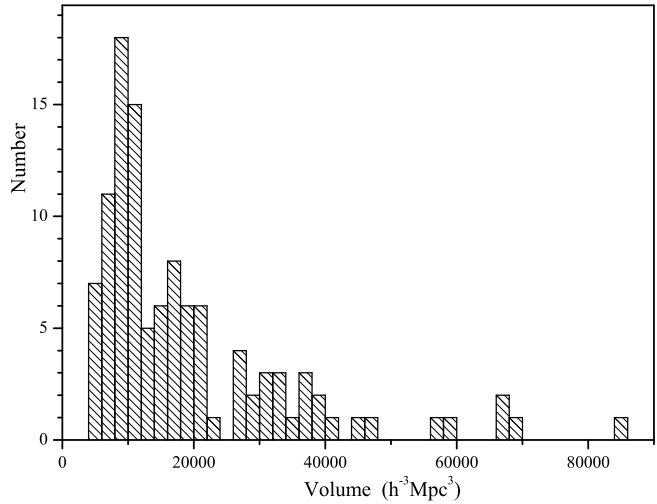}
} \figcaption{Distribution (histogram) of void volumes with
$R_{seed}> 9h^{-1}$~Mpc. }
\end{figure}

\begin{figure}
\centerline{%\psfig{figure=f1.eps,height=10cm,angle=0}
\includegraphics[]{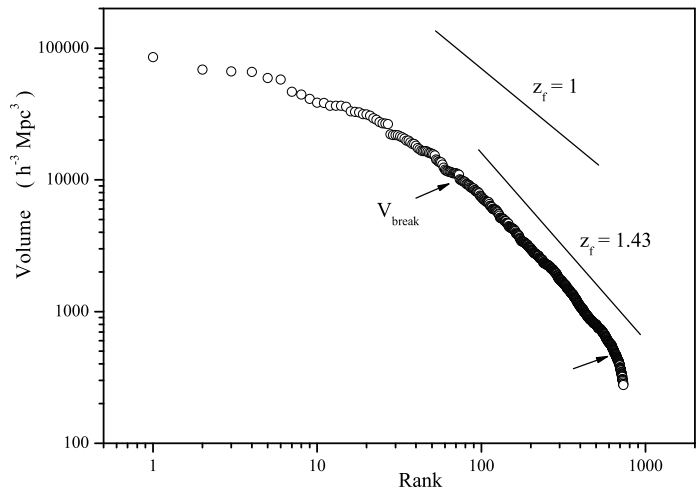}
} \figcaption{Void volume vs. void rank. $V_{break}$ corresponds
to the break in the dependence. The arrows indicate the range of
the linear fit. }
\end{figure}

\begin{figure}
\centerline{%\psfig{figure=f1.eps,height=10cm,angle=0}
\includegraphics[]{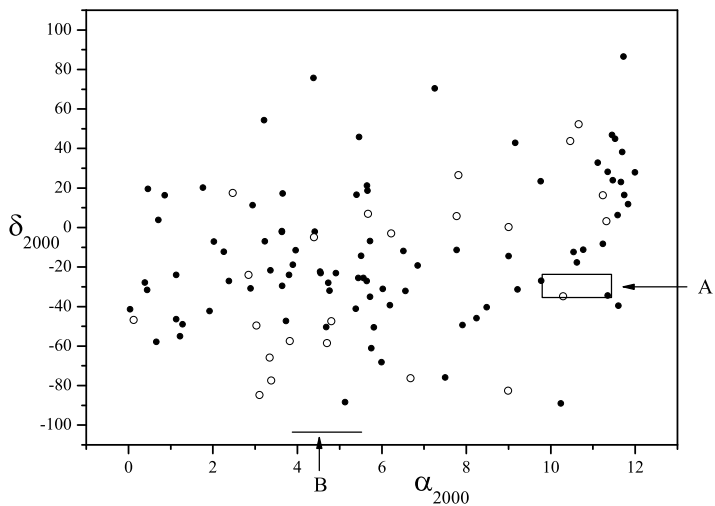}
} \figcaption{Directions of the greatest elongations of voids with
$R_{seed}> 9h^{-1}$~Mpc (transferred to the hemisphere $0^h <
\alpha < 12^h$). The open circles indicate the directions of 23
voids that do not touch the sample boundaries; the filled circles
indicate the directions of the voids that touch the sample
boundaries. A -- the rectangle of line-of-sight directions for the
2dFGRS sample used here; B -- the $\alpha$ range of directions
perpendicular to the line of sight. }
\end{figure}

\begin{figure}
\centerline{%\psfig{figure=f1.eps,height=10cm,angle=0}
\includegraphics[]{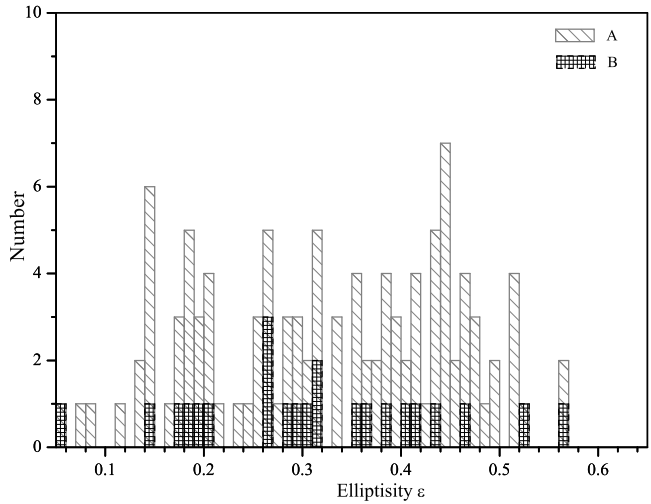}
} \figcaption{Distribution (histogram) of voids ellipticities: A
-- all voids with $R_{seed} > 9.0 h^{-1}$~Mpc and B -- voids with
$R_{seed} > 9.0 h^{-1}$~Mpc that do not touch the boundaries. }
\end{figure}

\begin{figure}
\centerline{%\psfig{figure=f1.eps,height=10cm,angle=0}
\includegraphics[]{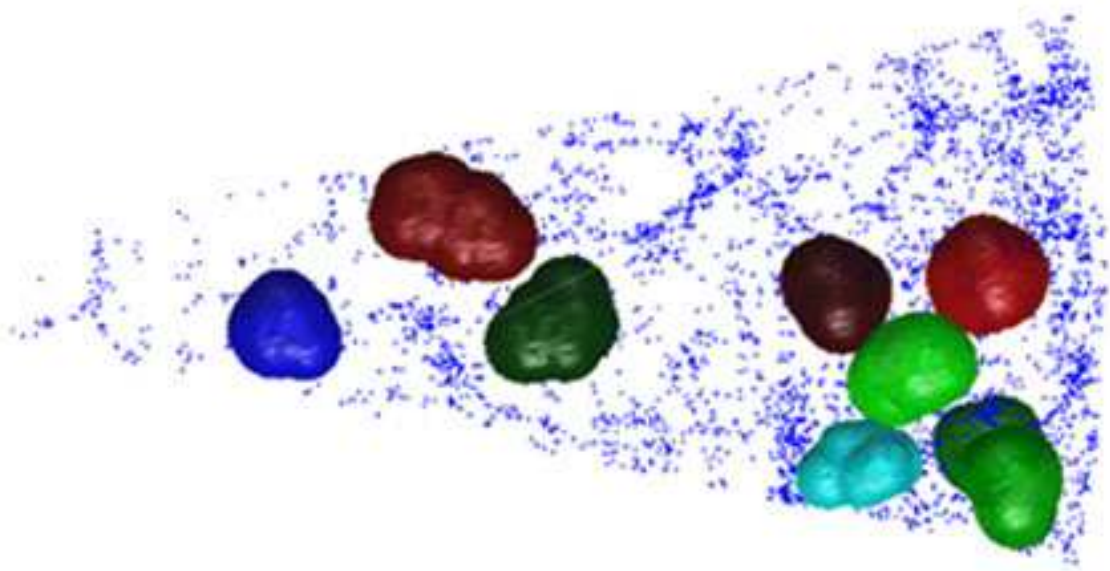}
}
\figcaption{Nine largest identified voids and galaxies of the
2dFGRS work sample with $-32^\circ < \delta < -27^\circ$. }
\end{figure}
%\fbox{ \psfig{width=10cm, bb=0in 0in 7.61in 6.36in,
%file=fig8.eps}}

\end{document}